\documentclass[12pt]{article}
\usepackage{epsfig}

\newcommand{\fup}[1]{{\small\raisebox{1ex}{#1}}}
\newcommand{\omegacp}{\omega^{\scriptstyle \mathrm {CP}}}
\newcommand{\eigveccp}{\mathbf {e}^{\scriptstyle \mathrm {CP}}}
\newcommand{\omegavff}{\omega^{\scriptstyle \mathrm {VFF}}}
\newcommand{\eigvecvff}{\mathbf {e}^{\scriptstyle \mathrm {VFF}}}

\begin{document}

\begin{center}
{\large \bf %
Vibrational properties of a sodium tetrasilicate  glass: 
{\it Ab initio} versus Classical Force Fields
}\\
\end{center}

\begin{center}
{%
Simona Ispas\footnote{Author to whom correspondence should be addressed:
e-mail: simona.ispas@ldv.univ-montp2.fr, Tel. ++ 33(0)467144979, 
Fax ++ 33(0)467143498}\fup{1},  Nikolay Zotov\fup{2}, 
 Sidoine  De Wispelaere\fup{1}, 
 and Walter  Kob\fup{1}
}
\end{center}

\begin{center}
{
	 \fup{1} {Laboratoire des Verres, Universit\'e Montpellier 2,}\\[1ex]
{Place E. Bataillon, 34095 Montpellier Cedex 5, France}\\
\fup{2} {Mineralogisch-Petrologisches Institut,}\\
{ University Bonn, Poppelsdorfer 
Schloss,}\\
{\small D-53115 Bonn, Germany}, 
}\\

\date{\today}
\end{center}

\begin{abstract}
 We have determined the vibrational properties  of a  sodium  tetrasilicate 
  (Na$_{2}$Si$_{4}$O$_{9}$)  glass model generated by molecular 
  dynamics simulations. 
  The  study  has been  carried out  using a classical valence force 
  fields approach as well as an 
  {\it ab initio } approach in the framework of the density functional theory. 
   The  total and partial vibrational densities of states (VDOS) are presented, 
   as well as  some characteristics 
   of the vibrational modes (participation ratios, correlation lengths). 
   For the low-frequency bands
   below 500 cm${}^{-1}$, we find that the shapes of the two calculated VDOS as well as those of their corresponding partial VDOS
   are quite similar.  
    For the intermediate- and high-frequency ranges, 
   we observe larger discrepancies between the two calculations. 
   Using the eigenmodes of the dynamical matrix we also calculate the
   polarized Raman spectra  within the 
bond-polarizability approximation. 
 We find an overall agreement between the calculated 
parallel  polarized (VV) Raman spectra and
the corresponding experimental spectrum.
Regarding  the perpendicular  depolarized  (VH) Raman spectrum, the
comparison of the calculated spectra to the experimental data indicates 
a need for an  adjustment of
the VH bond-polarizability parameters.
\end{abstract}


\newpage

\section{Introduction}
\label{sec:intro}

The nature of the vibrational excitations of silicate  glasses and melts represents a 
challenging problem of  condensed matter physics, glass and earth sciences. 
Significant experimental and theoretical efforts have been made in order to understand 
the vibrational properties of these materials, because they provide valuable information
about the underlying microscopic structure and other anomalous physical properties of 
silicate glasses \cite{Courtens_SolStComm01}. However, the lack of long-range translational order considerably 
complicates this task and requires  new approaches for measurement, analysis and simulations 
of the vibrational spectra of glasses and melts as well as 
a careful investigation on how the vibrational
spectra depend on the details of the considered glass 
 (chemical composition, thermal history, etc.).

Therefore we compare  in the present paper the  vibrational density of states (VDOS) and several 
vibrational characteristics of an {\it ab initio} 
model of sodium tetrasilicate glass Na$_2$Si$_4$O$_9$ (denoted hereafter NS4) calculated by 
two different methods.  The first calculation is based on an {\it ab initio} treatment 
of the interatomic forces and  is performed in the framework of the density functional theory, 
while the second one is performed in the approximation of 
a classical  valence
force fields (VFF) potential which is especially suited for describing the dynamics of partially 
covalent materials like silicate glasses.

The structure of the NS4 glass has been  intensively investigated by different experimental 
methods \cite{NS4_expdata}, reverse Monte Carlo (RMC) \cite{RMC_Zotov_PhysChemMin98}
 as well as  classical molecular dynamics (MD) simulations \cite{NS4_MDcalc,RamanNS4-Zotov_PRB99}. 
 Recently, the structural and  electronic properties of the NS4 glass 
have been modeled by a combined Car-Parrinello and classical MD  simulation
\cite{CP_Ispas_PRB01,CP_Ispas_JNCS02}. The attention given to this system is justified by the 
fact that it can be used as a prototype for more complicated aluminosilicate 
and hydrous silicate glasses.

The aims of the present study are: {\it (i)} to test  the quality of the {\it ab initio} NS4 
vibrational dynamics by comparing  the calculated and experimental polarized Raman spectra;
{\it (ii)} to test  the quality and the transferability of a previously proposed 
\cite{RamanNS4-Zotov_PRB99} valence force fields potential 
for sodium silicate glasses by comparison 
with {\it ab initio} calculations; and  {\it (iii)} 
to analyse  the effects of adding Na$_2$O to a-SiO$_2$ 
on the character of  the vibrational characteristics of the glass.
We would like to note that the present {\it ab initio} 
calculations using the Car-Parrinello (CP) method involve  various approximations
(pseudopotentials, truncated plane-wave basis, etc.)
Nevertheless we may expect
 that the CP approach leads to a better estimation of the
 real interatomic forces by comparison to the VFF approach. On the other hand,
the computation time needed to perform a VFF type calculation is
generally very small compared to the one used in  first principles calculations.


\section{Calculation procedures}
\label{sec:procedures}

\subsection{The glass models}
\label{sec:glassmodel}

The  amorphous NS4 model used in this study contains  90 atoms (24 silicon atoms, 54 oxygen atoms and 12 sodium atoms)
confined in a cubic box of edge length 10.81 \AA \ corresponding to the experimental NS4 mass density of 2.38 g/cm${}^3$ \cite{bansal_doremus}.
The glass model has been generated using  combined classical and {\it ab initio} molecular dynamics (MD)
 simulations. 
 For this we have firstly performed  the liquid
 equilibration, the quench, and the initial part of
  the low temperature relaxation of the system 
  by using a classical force field. Subsequently 
  we have refined the obtained structure   within the framework of 
 the {\it ab initio} MD. We note that this approach
 was successfully employed for the study of the structural, electronic and vibrational properties of
 vitreous SiO${}_2$ \cite{SiO2_Benoit_EPJB00,SiO2_Benoit_EuroPhysLet02}, as well as for a 
 structural and electronic study of vitreous NS4 
 \cite{CP_Ispas_PRB01,CP_Ispas_JNCS02}. 
 
 The classical MD simulations 
were performed with an interatomic potential which is  a modification 
of the  potential proposed by  van Beest et al. \cite{bks_sio2},  derived in order to study sodium 
silicates \cite{Horbach-bks-na}. The NS4 model was generated by quenching a well equilibrated liquid at
3500 K to 300 K, with a quench rate equal to $5\cdot 10^{13}$ K/s. 
The glass model obtained in this way  was
annealed at 300 K for 70 ps. The final atomic coordinates and velocities after
 the classical relaxation, were used as initial coordinates and velocities for a short  ($\approx 0.5$ ps) {\it ab initio}
MD simulation.
The  {\it ab initio} simulations were performed in the framework  of the
  Car-Parrinello (CP)  method  \cite{cp_85,marx-hutter_00} 
 using the CPMD software \cite{CPMD-code}.  

 In the first principles calculations, 
the electronic structure calculations were treated via
 the Kohn-Sham formulation \cite{Kohn-Sham} of density functional theory, 
 within the generalized gradient
approximation  employing the B-LYP functional \cite{BLYP}.
 The valence Kohn-Sham orbitals were expanded
in a plane-wave basis set defined by an energy cutoff of 70 Ry, 
 at the $\Gamma$-point of the supercell. 
 The core-valence interactions were
 described by norm-conserving pseudopotentials of 
 the Bachelet-Hamann-Schl\"uter type for Si atoms \cite{BHS-pseudo} and
 of the Trouiller-Martins type for O atoms \cite{TrouillerMartins-pseudo}.
Further details are given in Ref. \cite{CP_Ispas_PRB01}. 
The CP dynamics was performed using a fictitious electronic mass 
of 800 a.u. and a time step of 0.097 fs.
In the present CP calculations, we used  for the sodium atoms  
a semi-core Trouiller-Martins 
norm-conserving pseudopotential \cite{TrouillerMartins-pseudo} 
instead of the Goedecker semi-core type used in Ref. \cite{CP_Ispas_PRB01}.
This change allowed a better estimation of the internal pressure of the model. 
Nevertheless, the same structural modifications as reported in 
\cite{CP_Ispas_PRB01} occurred immediately after the CP dynamics 
was switched on, i.e. the shortening  of the Si-NBO bond and 
the lengthening of the Si-BO and Na-NBO bonds (in the remainder of the text,
 'BO' denotes the bridging oxygens, while 'NBO' denotes 
 the non-bridging oxygens).
The CP run leads to structural  relaxation without  changes 
in the network topology, the ring and Q-species distributions
 being determined in  the classical stage. 

In order to produce the first NS4 glass model
 (called the {\it ab initio} NS4 model) 
 used further for the vibrational calculations, we relaxed  to $0$ K the
 glass model obtained at the end of the CP simulation described above.
The second glass model was produced by performing a relaxation of the 
{\it ab initio} NS4 model with  a harmonic valence 
force field (VFF) potential of the  Kirkwood-type  
(see Ref. \cite{RamanNS4-Zotov_PRB99} for further details as well as the next 
subsection where we give the values of  the VFF potential parameters
used in this work). 
At the end of the VFF relaxation, the NS4 model had practically 
the same average Si-O and Na-NBO bond lengths as the 
starting {\it ab initio} model.
However we found  a small decrease ($\approx 3\% 
$) of the Na-BO bond length
 as well as a  decrease of the standard deviation of 
the average Si-O bond length distribution, 
from 0.033 \AA\, to 0.019 \AA.
 In addition we  observed a decrease of the standard deviation 
 of  the intratetrahedral O-Si-O angle distribution 
 (from 3.7 degrees to 2.2 degrees).

\subsection{VDOS and Raman spectra calculations}
\label{sec:VDOScomputing}

Within the first principles 
approach, the dynamical matrix 
was computed numerically by evaluating the second derivatives 
of the total energy with respect to atomic displacements 
($\approx 5.3 \cdot 10^{-3} $ \AA \, for each atom), 
which were calculated using  
the finite differences of the atomic forces.  
(We note that, for these latter calculations as well as for the 
{\it ab initio} relaxation to $0$ K mentioned in the previous subsection,
we used an energy cutoff of 90 Ry for the plane-wave basis
 expansions of the Kohn-Sham orbitals.)
The direct diagonalization of the 
dynamical matrix provides the
 eigenvalues  $\{\omegacp_p\}$ and   the corresponding normalized eigenvectors 
 $\{{\eigveccp} (\omegacp_p)\}$, $p=1, \ldots \, , 3N$, where $N=90$ is the number of  atoms in the model. 
Each eigenvector is a $3N$-dimensional vector which components  are proportional to the displacements of  the atoms in mode $p$.

 Within the VFF 
approach,  the dynamical matrix was calculated analytically and was 
diagonalized using the Householder method. 
The  detailed expressions of  the dynamical matrix elements are
given elsewhere \cite{Michailova_JNCS94}. The obtained vibrational 
frequencies will be denoted  $\{\omegavff_p\}$ and 
  the corresponding eigenvectors 
$\{{\eigvecvff}(\omegavff_p)\}$, $p=1, \ldots \, , 3N$.
In the present work we  used the
 following values for  the stretching force constants $\alpha_{ij}$~:
$\alpha _{\scriptstyle\mathrm{Si-BO}}=465 \, \mathrm{N/m}$,
 $ \alpha _{\scriptstyle\mathrm{Si-NBO}}=655 \, \mathrm{N/m}$,
$\alpha _{\scriptstyle\mathrm {Na-BO}}=25 \, \mathrm{N/m}$, and
$\alpha _{\scriptstyle\mathrm{Na-NBO}}=30 \, \mathrm{N/m}$, while the bending  force constants $\beta_{ijk}$ were  
$\beta_{\scriptstyle\mathrm{BO-Si-BO}}=\beta_{\scriptstyle\mathrm{BO-Si-NBO}}=
\beta_{\scriptstyle\mathrm{NBO-Si-NBO}}=35\, \mathrm{N/m}$ and 
$\beta_{\scriptstyle\mathrm{Si-BO-Si}} =14 \, \mathrm{N/m}$. 
We note that we have used larger Na-O stretching and 
slightly larger Si-BO-Si bending force constants than
the values reported in  Ref. \cite{RamanNS4-Zotov_PRB99} since
these values led to a better agreement between 
the calculated and experimental heat capacities for sodium silicate glasses 
\cite{HC_Zotov_JNCS2002}. 

The knowledge of all the eigenmodes allows us to calculate the
 reduced polarized Raman spectra of the NS4 glass model using the bond 
 polarizability approximation (BPA) \cite{Alben_PRB75} presented in details 
in Refs. \cite{RamanNS4-Zotov_PRB99,Bell-Hibbins1975}.
The BPA parameters used in the present work are : 
$A'_{\scriptstyle\mathrm{Si-BO}}=0.5 \, \mathrm \AA^3$, 
$\gamma_{\scriptstyle\mathrm{Si-BO}}=0.05 \, \mathrm \AA^3$,
$A'_{\scriptstyle\mathrm{Si-NBO}}=1 \, \mathrm \AA^3$, 
$\gamma_{\scriptstyle\mathrm{Si-NBO}}=0.05 \, \mathrm \AA^3$,
$A'_{\scriptstyle\mathrm{Na-O}}=0.05 \, \mathrm \AA^3$, 
$\gamma_{\scriptstyle\mathrm{Na-O}}=0 \, \mathrm \AA^3$,
where $A'_{\alpha-\beta}$ is the derivative of 
the parallel bond polarizability parameter with respect to the length of the bond
between atoms $\alpha$ and $\beta$, while
$\gamma_{\alpha-\beta}$ is 
the  perpendicular bond polarizability parameter for the ${\alpha-\beta}$
bond ($\alpha,\, \beta$=Si,Na,BO,NBO,O).
 Recently, a comparison of the Raman 
scattering mechanism in $\alpha$-quartz using both {\it ab initio}
 and BPA methods
has shown that the BPA reproduces the {\it ab initio} 
Raman intensities within $15\%
$ \cite{UmariPasquarello_PRB01}. For amorphous silica models generated by
{\it ab initio} molecular dynamics simulations, the use of the BPA gives rise 
to Raman spectra in good agreement with experimental data \cite{Rahmani_Benoit03}.
This gives strong support for 
the application of the BPA for the calculation of the Raman spectra of 
 large models of silicate glasses.
The comparison of the Raman spectra 
calculated independently from the CP and the VFF eigenmodes will allow us 
to investigate  separately the effects of
 the BPA parameters and the eigenmodes on the accuracy of 
the Raman spectra calculations.


\section{Results}
\label{sec:results}

\begin{center}
\begin{figure}
\epsfxsize=420pt{\epsffile{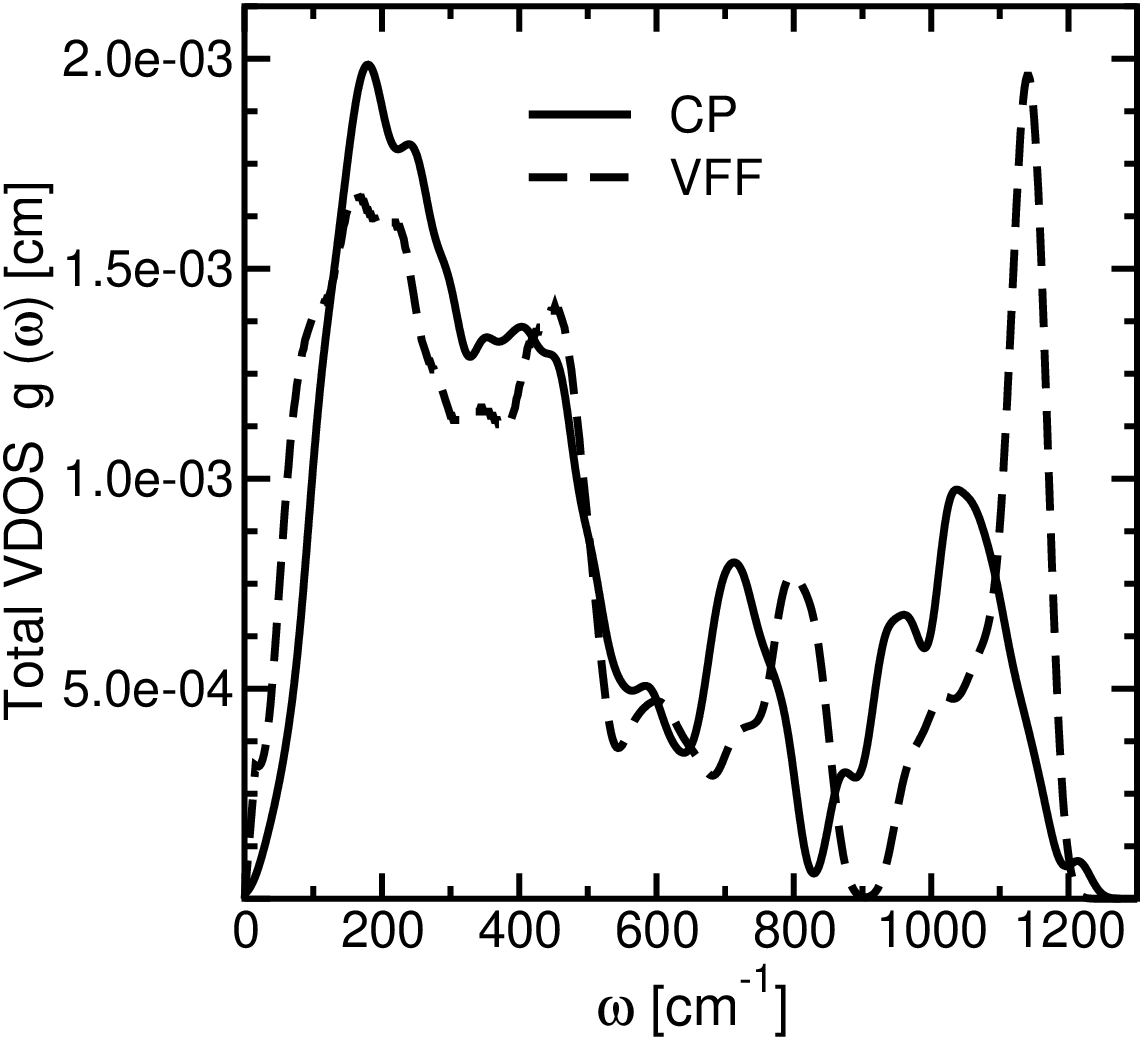}}
\caption{ }
\label{fig:VDOS-tot}
\end{figure}
\end{center}

In Fig.  \protect\ref{fig:VDOS-tot} we compare the CP and VFF vibrational densities of states. 
 Both VDOS are normalized to one, and we have used  the same uniform  
Gaussian broadening of full width at half maximum 
$2\sigma =40 \, \mathrm{cm}^{-1}$.  In both spectra we can
distinguish three ranges: a low-frequency range
 ($0-500 \,\mathrm{cm}^{-1}$), an intermediate frequency range
 ($500 - 900 \, \mathrm{cm}^{-1}$), and a high-frequency range
 ($900 - 1200 \, \mathrm{cm}^{-1}$ ). The
 low-frequency bands  arise generally from Na-O stretching and bending as well as from  Si-O bending and 
rocking motions.
 The strong mid-frequency band near $800 \, \mathrm{cm}^{-1}$ arises 
from complex motion with large displacements of the
 Si atoms against the BO atoms 
 \cite{RamanNS4-Zotov_PRB99,Galeener-Geissberger1983}.
The high-frequency modes correspond  mainly to Si-BO 
and Si-NBO stretching motions \cite {RamanNS4-Zotov_PRB99}. 

In spite of the fact that the intensities  of the CP 
low-frequency bands are slightly higher than the VFF ones,
 both approaches yield almost identical band positions 
and shapes. Larger discrepancies between the two VDOS are found in 
the mid- and the high-frequency ranges since the VFF mid-frequency 
bands are shifted to higher frequencies 
compared to the CP bands.
 Nevertheless the two approaches yield practically
 the same intensities (same number of vibrational modes). In the high-frequency range the shape of the
vibrational bands is generally the same but the VFF bands are
 shifted to higher frequencies, have larger intensities and are more narrow. 
 The latter effect may be attributed to the decrease of the standard
 deviation of the Si-O bond length distribution.

\begin{center}
\begin{figure}
\epsfxsize=400pt{\epsffile{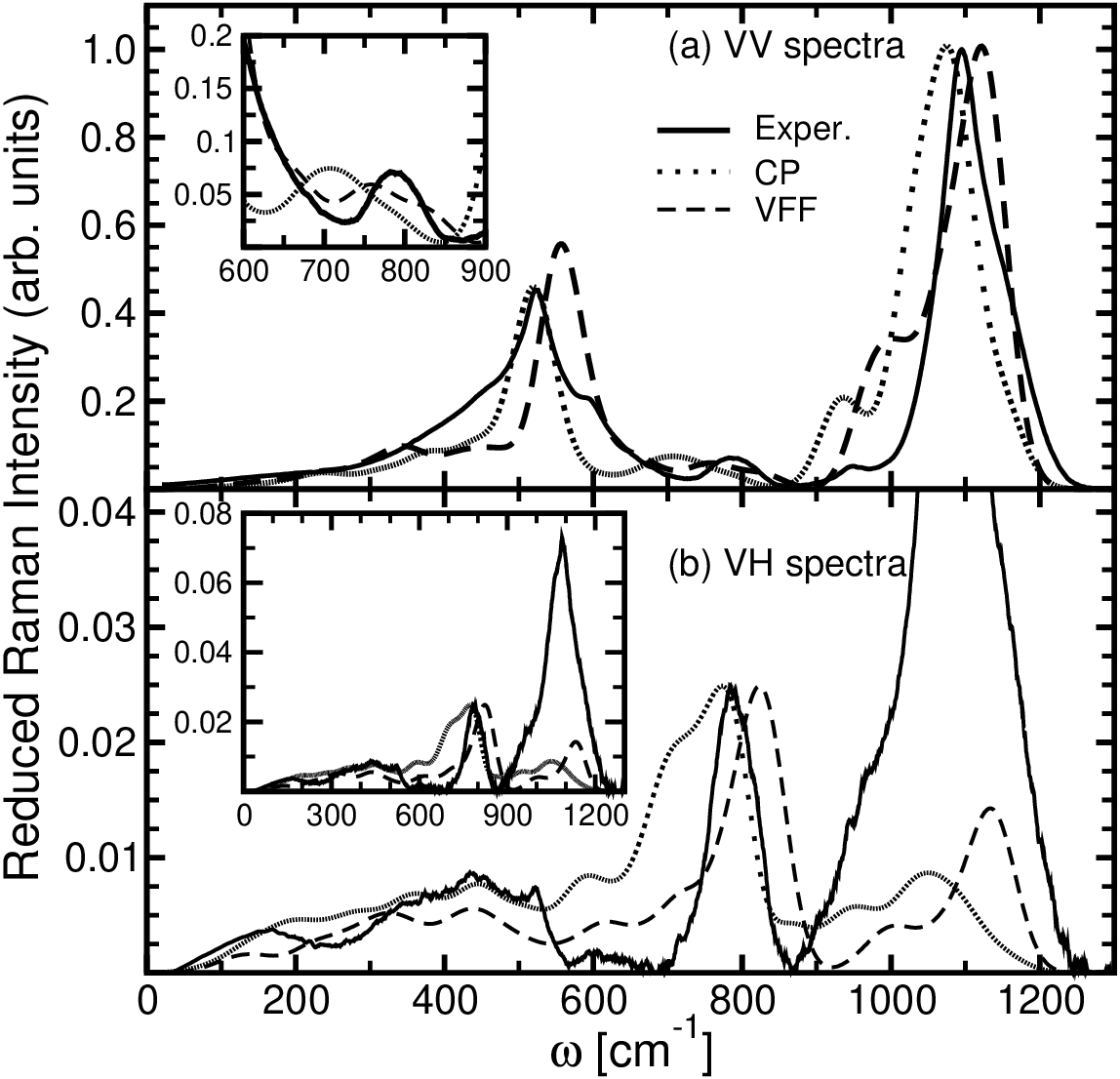}}

\caption{  }
\label{fig:Raman}
\end{figure}
\end{center}

Unfortunately, to the best of our knowledge, there are no 
VDOS data for the NS4 glass measured by inelastic neutron scattering to
 compare with our calculations. 
Therefore we compare in Fig. \protect\ref{fig:Raman} 
 the calculated and experimentally reduced polarized (VV and VH) Raman spectra.
The polarization is VV if the electric fields of the incident and the scattered light are parallel and is VH when they are perpendicular.
 We recall that the
 so-called reduced spectrum is obtained by multiplying the experimental 
measured spectrum with  the
 correction factor $\omega/[n(\omega)\cdot (\omega-\omega_0)^4]$ in order to have a
 temperature independent
 quantity ($n(\omega) $ is the Bose factor). 
The polarized Raman
spectra were measured on doubly-polished 
glass plates in backscattering geometry using the $514.5 \, \mathrm{nm}$ line
 of an Ar$^+$ laser and a  XY triple spectrometer equipped with 
a liquid-nitrogen cooled CCD detector  and confocal entrance optics  with 
$300 \, \mathrm s $ accumulation time per spectral window. 
The CCD spectra were binned with a step of $ 4\,  \mathrm{cm}^{-1}$.
 For the  calculations of the theoretical spectra, 
 the final spectra have been 
obtained by applying a uniform Gaussian broadening  
with full width at half maximum  equal to $30\, \mathrm{cm}^{-1}$ 
to the calculated Raman intensities. 
It should be noted that generally it is not possible to determine
experimentally absolute  Raman intensities.
 Therefore the VV spectra (Fig. \protect\ref{fig:Raman}a) were
  normalized so that the
 strongest peaks (at about $ 1100 \, \mathrm{cm}^{-1}$, 
 and characterized by a small depolarization
 ratio $I_{VH}/I_{VV}\approx 0.07$) 
 have the same maximum unit intensity, while the 
 VH spectra (Fig. \protect\ref{fig:Raman}b), were 
 normalized so that the strongest depolarized  peaks at 
 $\approx 800 \, \mathrm{cm}^{-1}$
 have the same intensity (with a depolarization ratio 
  $I_{VH}/I_{VV}\approx 0.35$).

In the VV Raman spectra (Fig. \protect\ref{fig:Raman}a) we distinguish also three bands. The previous
calculation  \cite{RamanNS4-Zotov_PRB99} of the VV Raman  spectra on NS4 glass models have assigned the 
low-frequency band at about $ 520 \, \mathrm{cm}^{-1}$ to symmetric 
 bending  of the Si-BO-Si linkages, the
mid-frequency band at about $ 780\, \mathrm{cm}^{-1}$ to the Si motions against the BO atoms
 and the  high-frequency band at about
 $1100 \, \mathrm{cm}^{-1}$ with a shoulder at about $920 \, \mathrm{cm}^{-1}$ 
 to the Si-NBO stretching motions. 

For the position and the relative intensity  of the low-frequency band, 
a good overall agreement between the  experimental  and the  CP calculated curve is observed although, e.g.,  
the width of the CP band is smaller and the shoulder in the experimental data at about
$ 600 \, \mathrm{cm}^{-1}$ is not reproduced. The agreement between 
the low-frequency VFF band and the experimental 
 spectrum is less good - the calculated  VFF intensity is stronger 
and the maximum peak position is shifted to slightly higher  frequencies. 
In other words, the VFF potential
overestimates the amplitudes of the Si-BO-Si
 symmetric bending vibrations.

The intensity of  the CP calculated mid-frequency band 
is in good agreement with the experiment but its position is shifted approximately
 $80 \,\mathrm{cm}^{-1}$
to lower frequencies.  The shape of the VFF corresponding band is similar to the experimental one,
but we note a smaller intensity and a slight shift to lower frequencies
 (see the inset in 
Fig. \protect\ref{fig:Raman}a).
  
Concerning the high-frequency band, we note that the position of 
the main peak is shifted approximately $ 20 \, \mathrm{cm}^{-1}$ 
downwards in the CP case and 
approximately $20 \, \mathrm{cm}^{-1}$ upwards in the VFF case indicating 
slightly weaker and slightly stronger Si-NBO interactions, respectively. 
 Both approaches yield much stronger intensity for the 
$920 \, \mathrm{cm}^{-1}$ band which arises mainly 
from Si-NBO stretching vibrations in SiO$_4$ tetrahedra with 
2 NBO per Si ($Q^{2}$ species) \cite{RamanNS4-Zotov_PRB99}.  

In the depolarized (VH) spectra, the calculated low-frequency bands 
  seem to be  correctly positioned (see Fig. \protect\ref{fig:Raman}b, 
  where we have lopped off the top of the VH spectra in order to see the
 details of the spectra in the low- and mid-frequency ranges). 
 Regarding the relative intensities, we note a good agreement for the CP band.
 In the mid-frequency range, the CP band is broadened and 
 shifted to slightly lower 
frequencies, while the VFF  band 
is shifted to slightly higher frequencies. In a previous 
calculation of the VH Raman spectra using a NS4 model generated by 
RMC simulations \cite{RamanNS4-Zotov_PRB99}, the agreement was better in this 
frequency range.  Since the mid-frequency band 
arises mainly from Si vibrations  in $Q^{4}-$species
\cite{RamanNS4-Zotov_PRB99}, the differences between the present VFF 
calculation and the results presented in Ref. \cite{RamanNS4-Zotov_PRB99} 
are to be attributed to the different degree  of polymerization 
of the corresponding models \cite{RamanNS4-Zotov_PRB99,CP_Ispas_PRB01}.

 In the high-frequency range (see the inset in Fig.\protect\ref{fig:Raman}b) the calculated 
depolarized Raman intensities are much weaker  in {\it both} 
calculated spectra which indicates that the corresponding Si-O bending BPA 
parameters are too small.


\section{Discussion}

\label{sec:discussion}

\begin{center}
\begin{figure}
\epsfxsize=400pt{\epsffile{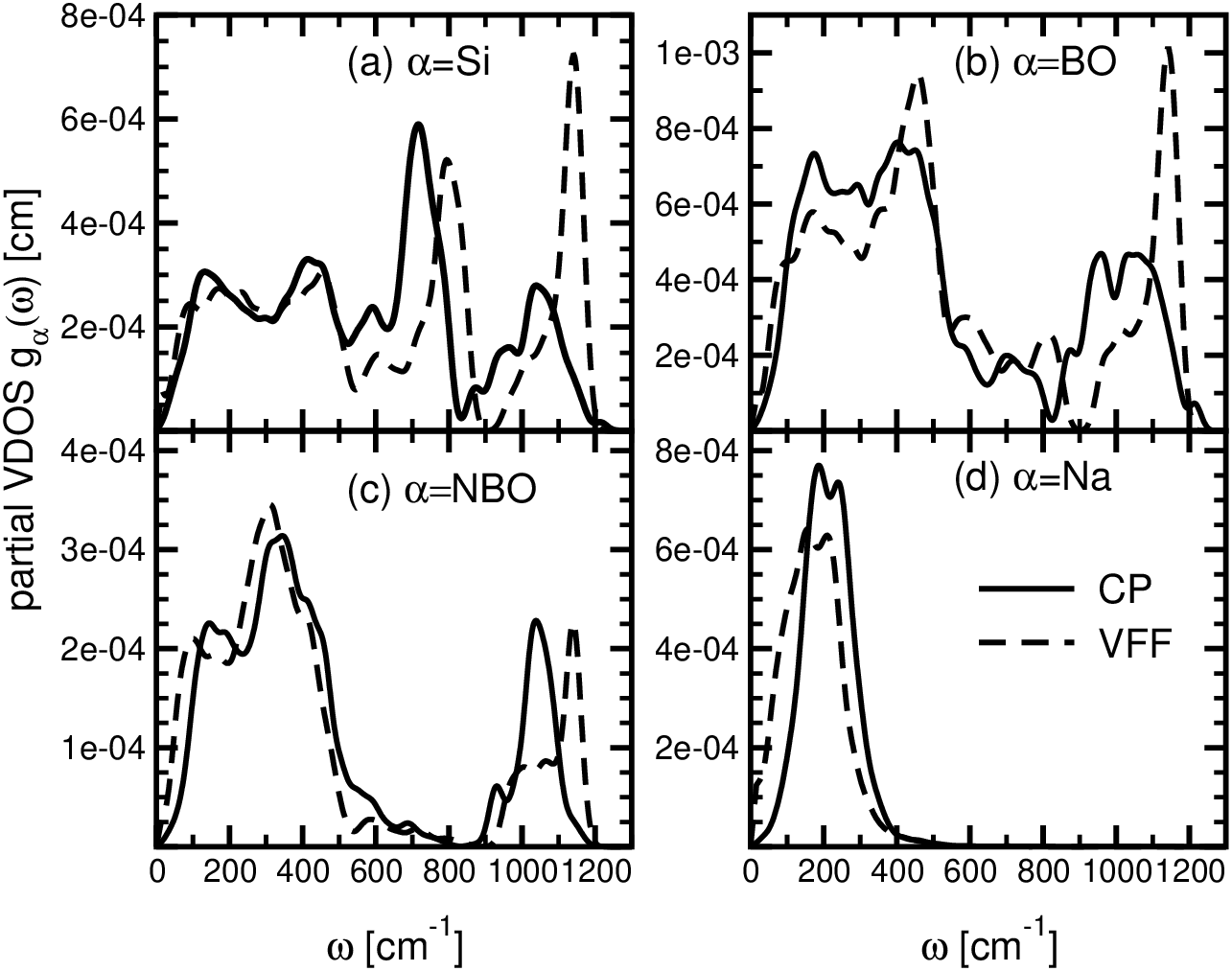}}

\caption{ }
\label{fig:partialvdos}

\end{figure}
\end{center}

In order to understand  
the differences  in the VDOS and Raman spectra computed  
within the CP and VFF approaches, we have analysed first the 
corresponding partial VDOS.  The partial VDOS curves $g_\alpha (\omega)$ in 
Fig. \protect\ref{fig:partialvdos} 
 ($\alpha =$ Si, BO, NBO, Na) have been defined as~:
\begin{eqnarray}
g_\alpha (\omega_p)=  g(\omega_p)\,\sum_{i \in \alpha} | 
{\mathbf e}_i (\omega_p)|^2
\end{eqnarray}
where ${\mathbf e}_i (\omega_p)$, $i=1, \,\ldots\,, N$, 
are 3-component real space eigenvectors.

 For the low-frequency bands, 
we have an overall good agreement between the CP and VFF partial 
VDOS. 
The band maxima positions for the Si (Fig. \protect\ref{fig:partialvdos}a), 
BO (Fig. \protect\ref{fig:partialvdos}b) and
 NBO (Fig. \protect\ref{fig:partialvdos}c) atoms are very close 
($\pm 30\, \mathrm{cm}^{-1}$). 
In this frequency range the
NBO atoms participate mainly in O-Si-O bending vibrations 
\cite{RamanNS4-Zotov_PRB99}. 
The biggest intensity differences are observed in the 
BO (Fig. \protect\ref{fig:partialvdos}b) and
Na (Fig. \protect\ref{fig:partialvdos}d) partial VDOS, which 
can explain the differences in the total VDOS 
in this frequency range.
 
As can be seen in 
Fig.  \protect\ref{fig:partialvdos}d, 
both the CP and VFF calculations predict a  dominant  Na 
contribution only in the low-frequency range, but the VFF 
peak is shifted ($\approx  40\, \mathrm{cm}^{-1}$) 
as a whole to  lower frequencies. 
From a  previous VFF analysis 
\cite{RamanNS4-Zotov_PRB99} of the vibrational 
characteristics of two NS4 structural models obtained by classical 
MD and RMC simulations, it 
followed that the Na motion dominated below $
\approx 100 \,\mathrm{cm}^{-1}$. In the present VFF calculation, 
the Na contribution reaches a maximum at 
$\approx 170\,  \mathrm{cm}^{-1}$. Thus the Na-O stretching force constants used in 
the present work describe better the Na-O motion than  the
force constants reported in Ref. \cite{RamanNS4-Zotov_PRB99}.

At intermediate frequencies,  the CP and the VFF partial
 VDOS  for the Si, BO and NBO atoms have qualitatively 
the same shape but the ones for Si and BO are shifted to 
lower frequencies in the CP calculation (Fig. \protect\ref{fig:partialvdos}a 
and b). 
These shifts are similar to the  observed shift
 of the mid-frequency band in the VV calculated using the CP 
eigenmodes. 
So it seems that  the CP approach does not describe 
properly  the vibrational dynamics of the Si and BO atoms in 
the mid-frequency range. 
Interestingly, the Si
 motions are enhanced in the CP calculation while the BO motions 
are enhanced in the VFF calculation.

In contrast to this, the partial VDOS show that in the high-frequency 
range all of the VFF  partial VDOS bands are stronger and more narrow 
(especially for the BO and Si atoms) than the CP ones. 
This feature may be related to the stronger localization of the CP modes
in this  frequency range, as it will be discussed  below.

\begin{center}
\begin{figure}
\epsfxsize=400pt{\epsffile{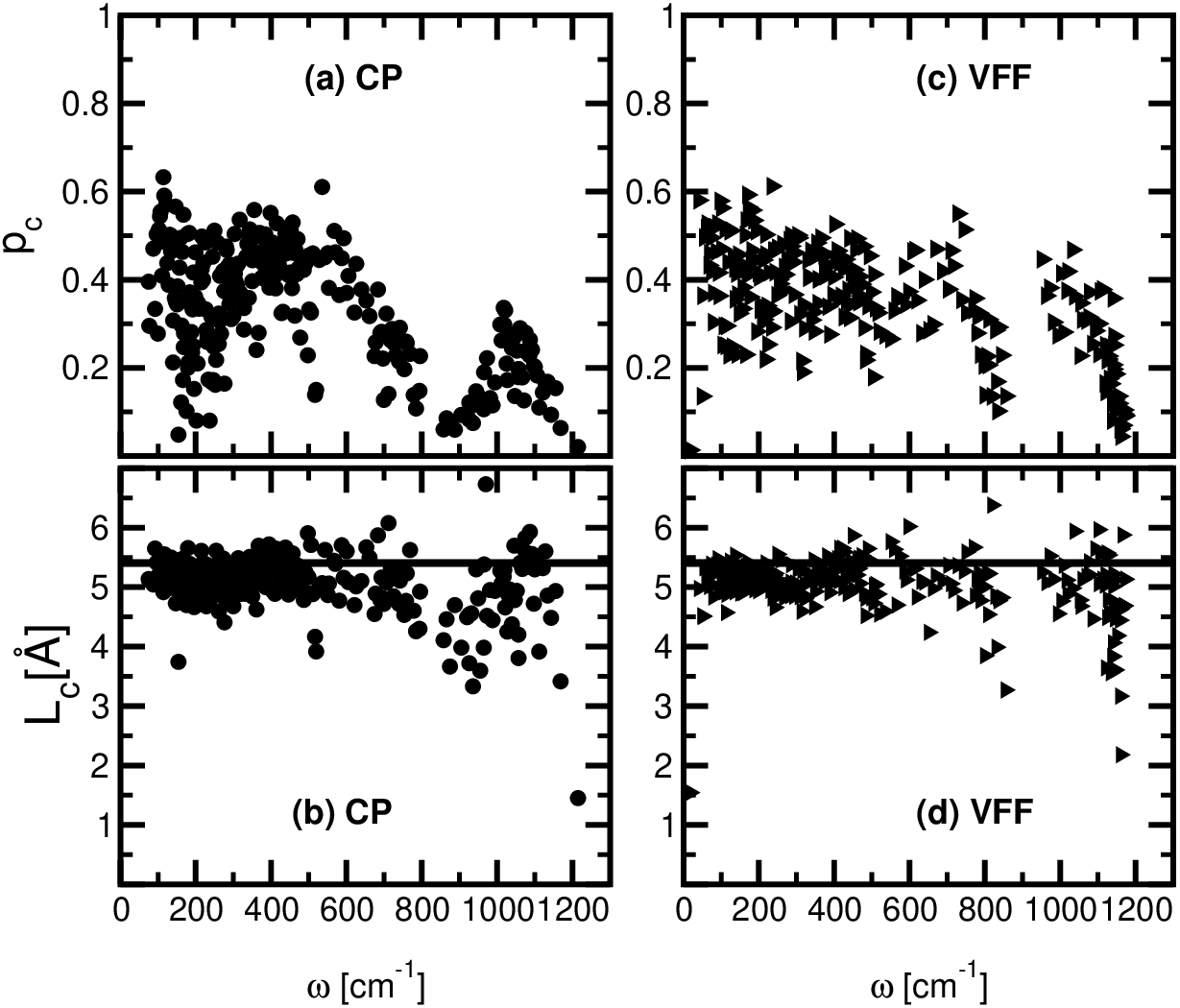}}

\caption{ }
\label{fig:pr-cl}
\end{figure}
\end{center}

As a first measure for  the localization of the modes  we consider  
 the participation ratio $p_c$ which, for a given eigenmode $p$, 
  is defined as follows \cite{Bell} :
\begin{eqnarray}
p_c(\omega_p)= \frac{(\sum_{i=1}^N |{\mathbf u}_{i}(\omega_p)|^2)^2}
{N\sum_{i=1}^N |{\mathbf u}_{i}(\omega_p)|^4},
\end{eqnarray}
where $\displaystyle {\mathbf
u}_i (\omega_p)={\mathbf e}_{i}(\omega_p)/\sqrt{M_i}$ is 
the atomic displacement of atom $i$ in
mode $p$ and $M_i$ is the mass of atom $i$.
The frequency dependences of the CP and VFF 
participation ratios are shown  in Figs. \protect\ref{fig:pr-cl}a and \protect\ref{fig:pr-cl}c. 
As can be seen from  Fig. \protect\ref{fig:pr-cl}a and \protect\ref{fig:pr-cl}c, the CP modes are 
generally more localized than the VFF ones, especially around 200 
and above $ 900  \, \mathrm{cm}^{-1}$. This is an interesting 
result which merits to be studied in more details.
 The localization is also enhanced near the edges 
of the three main VDOS bands  for both the CP and the VFF participation ratios. These are the so-called band tails containing   localized modes, which have  been already observed
for a larger NS4 glass models in Ref. \cite{RamanNS4-Zotov_PRB99}, 
as well as for silica glass models 
\cite{RamanNS4-Zotov_PRB99,SiO2_Benoit_EuroPhysLet02,TaraskinElliott97-prb}.
In the CP case, these tails are more pronounced, in particular at the top of the lower
 band and on both sides of the high-frequency bands. The analysis of the frequency 
dependences of the vibrational correlation  lengths $L_c(\omega_p)$ 
which are plotted in Fig. \protect\ref{fig:pr-cl}b and \protect\ref{fig:pr-cl}d, confirms the
 localization of the high-frequency modes. 
 The vibrational correlation length is defined 
  as:
 \begin{eqnarray}
L_c(\omega_p)= 
\sqrt{\frac{\sum_{i=1}^N |{\mathbf r}_i- {\mathbf r}_p|^2 |
{\mathbf u}_{i}(\omega_p)|^2}
{\sum_{i=1}^N |{\mathbf u}_{i}(\omega_p)|^2}
}
\end{eqnarray}
where ${\mathbf r}_i$ is the position of  atom $i$, ${\mathbf r}_p$ is
the center of mode $p$ displacements, defined as
$\displaystyle {\sum_{i=1}^N {\mathbf r}_i |{\mathbf u}_{i}(\omega_p)|^2}$. 
In this sense, the vibrational correlation length 
 \cite{MarinovRamanSi_PRB97,JinVashishtaSiO2_PRB93} is the standard deviation
of ${\mathbf r}_p$, and
gives a measure of the spatial localization of the vibrational modes by indicating that the amplitude of the atomic
vibrations decreases significantly beyond this  correlation length. 

\section{Conclusions}
\label{sec:conclusions}

In this work, we have investigated some aspects 
of the vibrational dynamics of   a model of  sodium tetrasilicate glass, 
constructed by combined classical and Car-Parrinello molecular dynamics 
simulations. Its vibrational dynamics has been then studied using
 two different approaches -
 a first principles one and a classical valence force fields one. 
 The valence force field serves as a natural approach to analyze 
  the character of the vibrational modes in terms of simple motions like 
  stretching, bending and rocking. 
 Within the bond polarization approximation,
 we have calculated the polarized Raman spectra  and compared with experimental Raman data.

 The VDOS calculated by the two methods have very similar shape, especially 
 for the low-frequency bands below $500 \, \mathrm{cm}^{-1}$.
In the high-frequency range, the VFF partial VDOS for the Si, 
BO and NBO atoms are shifted to higher frequencies and they are narrower.
This may indicate that the potential VFF parameters have to be 
adjusted in order to better reproduce the CP vibrational characteristics of our
NS4 model (an increasing number of low frequency modes, 
the position as well as the localization of the high frequency modes).

The comparison of the calculated and experimental VV Raman spectra
 shows that a good agreement between the relative intensities of the low-
and high-frequency bands is achieved using the CP eigenmodes. 
This indicates that the corresponding BPA parameters are approximately
 correct and can be used for 
calculation of the VV Raman spectra of  structural models of silicate 
glasses that are still too large to be fully treated {\it ab initio}. 
However, despite the fact that the CP eigenvectors are 
calculated from first principles they do not reproduce well the 
position of the mid-frequency VV and VH Raman bands at around  
$780 \, \mathrm{cm}^{-1}$. Additional improvement 
of the BPA parameters describing the high-frequency depolarized
 Raman bands  is also necessary.  

The analysis of the character of the vibrational modes shows that the 
localization of the CP vibrational modes  is stronger than the one of
the VFF modes, in particular  on the high $\omega $ side  of the low-frequency
band ($\approx 200\, \mathrm{cm}^{-1}$)
and on both sides of the high-frequency bands.

\section*{Acknowledgments}
We  thank  Magali Benoit 
for very interesting and stimulating 
discussions. Part of this was supported by the European 
Community's Human Potential Program under contract 
HPRN-CT-2002-00307, DYGLAGEMEM. 
The financial support from the German Science 
Foundation under project SFB 
408 for N. Zotov is kindly acknowledged.
The {\it ab initio} simulations have been 
performed on the IBM/SP3 at CINES, Montpellier, FRANCE.

\newpage

\newpage

\centerline{FIGURE CAPTIONS}

\noindent {\sc Figure 1~:} Vibrational densities of states of the
 {\it ab initio} NS4 glass model obtained within 
 the Car-Parrinello {\it ab initio}
  approach (solid line) and  within
 a classical valence forcefield approximation (dashed line).\\

\noindent{\sc Figure 2~:} Frequency dependence of the VV (a) and the 
VH (b)  Raman intensities calculated within the bond-polarizability 
approximation using 
the CP eigenmodes (dotted lines) and
 the classical VFF eigenmodes (dashed lines).
 The solid lines are the experimental spectra.\\

\noindent{\sc Figure 3~:} 
Partial VDOS for Si (a), BO (b), 
NBO  (c), and Na (d). 
 The solid and dashed
lines correspond to the CP and VFF calculations, respectively.
\\

\noindent{\sc Figure 4~:}  Frequency dependences of the participation $p_c$ 
and of the vibrational correlation length $L_c$ in the CP case 
 (panels a) and b) )
and in the VFF case (panels c) and d)).
The horizontal lines at a height equal to 5.405 \AA\, 
(i.e.  half of our simulation box length) are drawn as a guide for the eye in 
order to estimate the number of the normal modes with 
correlation length smaller or bigger than this value.

\newpage

\newpage

\newpage

\begin{thebibliography}{99}
\bibitem{Courtens_SolStComm01} 
E. Courtens, M. Foret, B. Hehlen, R. Vacher, 
Solid State Commun. {117} (2001) 187.

\bibitem{NS4_expdata} R. Dupree, D. Holland, P.W. McMillan, R.F. Pettifer,
 J. Non-Cryst. Sol. { 68}  (1984) 399;\\
 J. F. Emerson, P.E. Stallworth , P.J. Bray, 
 J. Non-Cryst. Sol. { 113} (1989) 253;\\
  G. H. Wolf, D.J. Durben, P. F. McMillan, 
  J. Chem. Phys.  { 93}  (1990) 2280; \\
 H. Maekawa, T. Maekawa, K. Kawamura, T. Yokokawa,  J.
Non-Cryst. Sol. { 127} (1991) 53; \\
J. K\"ummerlen, L. H. Merwin, A. Sebald, H. Keppler, 
J. Phys. Chem. { 96} (1992) 6405; \\
N. Zotov, H. Keppler, A.C. Hannon, A.K. Soper,
 J. Non-Cryst. Sol. { 202} (1996) 153.

\bibitem{RMC_Zotov_PhysChemMin98}
N. Zotov, H. Keppler,  
Phys. Chem. Minerals { 26} (1998) 107.

\bibitem{NS4_MDcalc} 
A. A. Tesar, A. K. Varshneya, 
J. Chem. Phys. { 87} (1987) 2986; \\
 C. Huang, A. N. Cormack,
  J. Chem. Phys. { 93} (1990) 8180 ;\\ 
C. Huang, A. N. Cormack, 
J. Chem. Phys. { 95}  (1991) 3634; \\
J. Oviedo, F. Sanz,  
Phys. Rev. B { 58} (1998) 9047; \\
E. Sunyer, P. Jund , R. Jullien, 
Phys. Rev. B { 65} (2002) 214203;\\
X. Yuan, A.N. Cormack,
J. Non-Cryst. Sol. {319} (2003) 31.

\bibitem{RamanNS4-Zotov_PRB99}
 N. Zotov, I. Ebbsj\"o, D. Timpel , H. Keppler, 
 Phys. Rev. B {60} (1999) 6383.

\bibitem{CP_Ispas_PRB01}
 S. Ispas, M. Benoit, P. Jund, R. Jullien,
 Phys. Rev. B {64} 
(2001) 214206.


\bibitem{CP_Ispas_JNCS02}
 S. Ispas, M. Benoit, P. Jund, R. Jullien, 
 J. Non. Cryst. Sol.  {307-310} (2002) 946.


\bibitem{bansal_doremus}
N. P. Bansal, R. H. Doremus, 
{\it Hanbook of Glass Properties}, Academic Press, INC., New York (1986).


\bibitem{SiO2_Benoit_EPJB00}
M. Benoit, S. Ispas, P. Jund, R. Jullien,
 Eur. Phys. J. B { 13} (2000) 631.

\bibitem{SiO2_Benoit_EuroPhysLet02}
M. Benoit, W. Kob, 
Eur. Phys. Lett. { 60} (2002) 269.




\bibitem{bks_sio2} 
B.W. H. van Best, G.J. Kramer, R.A. van Santen, 
Phys. Rev. Lett.  {64} (1990)    1955.


\bibitem{Horbach-bks-na}
J. Horbach, W. Kob, K. Binder, 
Chem. Geol. {174}  (2001) 87.

\bibitem{cp_85} 
R. Car, M. Parrinello, 
Phys. Rev. Lett. { 55}  (1985) 2471.

\bibitem{marx-hutter_00} 
D. Marx, J. Hutter, in
 {\it Modern Methods,
 Algorithms of Quantum Chemistry}, ed. J. Grotendorst, Forschungszentrum
  J\"ulich, NIC Series, {1} (2000) 301.

\bibitem{CPMD-code}
CPMD Version 3.3, J. Hutter, A. Alavi, T. Deutsch, M. Bernasconi, 
S. Goedecker, 
D. Marx, M. Tuckerman, M. Parrinello, 
MPI f\"ur Festk\"orperforschung , IBM Research (1995-1999).

\bibitem{Kohn-Sham}
 W. Kohn, L. Sham, 
 Phys. Rev. A 140 (1965) 1133.

\bibitem{BLYP} 
A.D. Becke,
Phys. Rev. A {38} (1988) 3098;\\
C. Lee, W. Yang, R.G. Parr,
Phys. Rev B {37} (1988) 785.


\bibitem{BHS-pseudo}
G.B. Bachelet, D.R. Hamann, M. Schl\"uter,
Phys. Rev. B {26} (1982) 4199.


\bibitem{TrouillerMartins-pseudo}
N. Trouiller, J.L Martins, 
Phys. Rev. B { 43} (1991)   1993.


\bibitem{HC_Zotov_JNCS2002} 
N. Zotov, 
J. Phys.: Condens. Matter {14} (2002) 11655.

\bibitem{Michailova_JNCS94} 
B. Michailova, N. Zotov, M. Marinov, L. Konstantinov, 
J. Non-Cryst. Sol. {168} (1994) 265.




\bibitem{Alben_PRB75} 
R. Alben, D. Weaire, J.E. Smith Jr., 
M.M. Brodsky, Phys. Rev. B { 11}  (1975) 2271.


\bibitem{Bell-Hibbins1975}
R.J. Bell, D.C. Hibbins-Butler,
J.Phys. C {9} (1976) 2955.


\bibitem{UmariPasquarello_PRB01} 
P. Umari, A. Pasquarello, A. Dal Corso, 
Phys. Rev. B { 63}  (2001) 094305.

\bibitem{Rahmani_Benoit03}
A. Rahmani, M. Benoit, C. Benoit,
arXiv: cond-mat/0306332, Phys. Rev. B (2003) in press.


\bibitem{Galeener-Geissberger1983}
F.L. Galeener, A.E. Geissberger,
Phys. Rev. B {27} (1983) 6199.

\bibitem{Bell} 
R. J. Bell, 
Methods Comput. Phys. {15} (1976) 215.


\bibitem{TaraskinElliott97-prb} 
S.N. Taraskin, S.R. Elliott,
 Phys. Rev. B {56} 
(1997) 8605.

\bibitem{MarinovRamanSi_PRB97} 
M. Marinov, N. Zotov, 
Phys. Rev. B, {55} 
(1997) 2938. 

\bibitem{JinVashishtaSiO2_PRB93} 
W. Jin, P. Vashishta, R.K. Kalia, J.P. Rino,
Phys. Rev. B, {48} 
(1993) 9359.  


\end{thebibliography}
\end{document}